\documentclass[11pt]{article}

\pdfoutput=1 
\usepackage[utf8x]{inputenc}
\usepackage{fontenc}
\usepackage{graphicx}
\usepackage{float}
\usepackage[lofdepth,lotdepth,caption=false]{subfig}
\usepackage{color}
\usepackage{amsmath,amssymb,latexsym}
\usepackage{cite}
\usepackage{booktabs}
\usepackage{tabularx}
\newcommand{\thb}{{\theta_b}}

\title{\bf{Probing Lorentz and CPT Violation in a Magnetized Iron Detector using Atmospheric Neutrinos}}
\author{Animesh Chatterjee$^1$\,\thanks{email: animesh@hri.res.in}~,~~Raj Gandhi$^1$\,\thanks{email: raj@hri.res.in}~,~~ Jyotsna Singh$^2$\,\thanks{email: jyo2210@yahoo.co.in}~,\\\\
\\
 {\it $^1$Harish-Chandra Research Institute, Chhatnag Road, Jhunsi, Allahabad 211 019, India }\\\\
 {\it $^2$University of Lucknow ,Lucknow 226007,India}\\\\
  }

\begin{document}

\maketitle
\renewcommand{\thefootnote}{\fnsymbol{footnote}}

\begin{abstract}

We study the sensitivity of the Iron Calorimeter (ICAL) at the India-Based Neutrino Observatory (INO) to Lorentz and CPT violation in the neutrino sector. Its ability to identify the charge of muons in addition to their direction and energy makes ICAL a useful tool in putting constraints on these fundamental symmetries. Using resolution, efficiencies, errors and uncertainties obtained from ICAL detector simulations, we determine sensitivities to $\delta{b}_{31}$, which parametizes the violations in the muon neutrino sector. We carry out calculations for three generic cases representing mixing in the CPT violating part of the hamiltonian, specifically , when the mixing is 1) small, 2) large, 3) the same as that in the PMNS matrix. We find that for both types of hierarchy, ICAL at INO should be sensitive to  $\delta{b}_{31}$ $\gtrsim$ $4\times10^{-23}$ GeV at 99 $\%$ C.L. for 500 kt-yr exposure, unless the mixing in the CPT violation sector is small.

\end{abstract}

\section{Introduction}
Invariance under the product of charge conjugation (C), parity (P) and time reversal (T), $\it{i.e}$ the CPT theorem ~\cite{W.Pauli,G.Grawert,Tsukerman:2010vx}, is a linchpin of present-day  quantum field theories underlying particle physics. It is noteworthy that this invariance under the product of purely discrete symmetries is actually a consequence of the invariance of the Lagrangian ($\cal L $) under  a connected continuous group, namely, proper Lorentz transformations. Additionally, it follows from the requirement that $\cal L $ be hermitian and the interactions in the underlying field theory be local, with the fields obeying the commutation relations dictated by the spin-statistics theorem. \\

Theories attempting to unify gravity and quantum mechanics may, however, break such seemingly solid pillars of low energy effective field theories (like the Standard Model (SM)) via new physics associated with the Planck scale. For a general mechanism for the breaking of Lorentz symmetry in string theories, see, for instance, \cite{Kostelecky:1988zi}. Other scenarios for such breaking have been discussed in \cite{G_Amelino_Camelia} and \cite{Douglas:2001ba}. Also, as shown in \cite{Greenberg:2002uu}, a violation of CPT always breaks Lorentz invariance, while the converse is not true. A framework for incorporating CPT and Lorentz violation into a general relativistically extended version of the  SM  has been formulated in \cite{Colladay:1998fq,Kostelecky:2003fs}. It is termed as the Standard Model Extension (SME), and our discussion in this paper will utilize the effective CPT violating (CPTV) terms that it introduces. Such terms, given the impressive agreement of the  (CPT and Lorentz invariant) SM with all present day experiments, must of necessity be small.

A characteristic attribute of neutrino oscillations is the amplification, via interference, of the effects  of certain small parameters ($\it{e.g.}$ neutrino masses) in the underlying SM lagrangian. As discussed in \cite{Kostelecky:2003xn}  and \cite{Kostelecky:2003cr}, the CPT violation manifest in an effective SME hamiltonian can also be rendered measurable in neutrino oscillation experiments. Its non-observation, on the other hand,  can be used to set impressive limits on CPT and Lorentz violation. This has been done, to cite a few  recent examples,  by IceCube \cite{Abbasi:2010kx}, Double Chooz \cite {Abe:2012gw}, LSND \cite {Auerbach:2005tq}, MiniBooNE \cite{PLB718}, MINOS \cite{Adamson} and Super Kamiokande \cite{Akiri}. Additionally, many authors have studied how best to parametrize and/or use neutrino oscillations and neutrino interactions to perform tests of CPT and Lorentz symmetry breaking in different contexts, ranging from neutrino factories and telescopes to long baseline, atmospheric, solar and reactor  experiments, including those looking for supernova neutrinos ~\cite{Coleman,Barger,Bahcall,Datta,amol,Diaz,JDiaz,Samanta,Engelhardt,SeanScully,JohnEllis,Int50,J335,IMotie,PGorham,SChakraborty,LSND,Brett,Zong,Enrico,Jorge,Rossi}.

In this article we study the sensitivity  of a large atmospheric magnetized iron calorimeter to CPTV  using a three flavour analysis with matter effects. As a reference detector for our study, we use the ICAL,  the proposed detector for  the India-Based Neutrino Observatory(INO)\cite{INO}.   The main physics objective of ICAL is the determination of the neutrino mass hierarchy through the observation of earth matter effects in atmospheric neutrinos, as discussed, for instance, in ~\cite{AGhosh,Blennow,Gandhi,Samanta1,Indumathi}. However, its lepton charge identification  capability renders it useful in measurements related to our purpose here.

 In what follows,  in Section 2 we review the perturbative phenomenological approach that allows us to introduce the effects of  CPT violation in the neutrino oscillation probability, based on  the SME. We examine effects at the probability level, in order to get a better understanding of the physics that drives the bounds we obtain using our event rate calculations. In section 3 we  describe our method of analysis, and  in section 4 we discuss our results, in the form the bounds on CPT violating terms. Section 5 summarizes our conclusions.

\section{ CPTV effects at the probability level}

 The effective Lagrangian for a single fermion field,  with Lorentz violation\cite{Colladay} induced by new physics at  higher energies can be written as
\begin{equation}
{\cal L} = i \bar{\psi} \partial_\mu \gamma^\mu \psi 
-m \bar{\psi} \psi            
- A_\mu \bar{\psi} \gamma^\mu \psi 
- B_\mu \bar{\psi} \gamma_5 \gamma^\mu \psi  \;,
\label{Lagran1}
\end{equation}
where $A_\mu$ and $B_\mu$ are real numbers, hence  $A_\mu$ and $B_\mu$  necessarily induce  Lorentz violation, being invariant under boosts and rotations, for instance. Such violation under the group of proper Lorentz transformations then leads to CPT violation \cite{Greenberg:2002uu}\footnote{CPT violation may also occur if particle and anti-particle masses are different. Such violation, however, also breaks the locality assumption of quantum field theories\cite{Greenberg:2002uu}. We do not consider this mode of CPT breaking in our work.}.

 The  effective contribution to the neutrino Lagrangian can then  be parametrized \cite{Coleman} as 
\begin{equation}
{\cal L}_\nu^{CPTV} =  
\bar{\nu}_L^\alpha \, b^{\alpha \beta}_\mu \, 
\gamma^\mu \, \nu_L^{\beta} \; 
\label{L-nu}
\end{equation} 
where $b_{\mu}^{\alpha \beta}$ are four Hermitian $3\times 3$
matrices corresponding to the four Dirac indices $\mu$ and $\alpha, \beta$ are flavor indices.
Therefore the effective Hamiltonian in  the vacuum for ultra-relativistic neutrinos with definite momentum p is 
\begin{equation}
{ H} \equiv \frac{{ M} { M}^\dagger}{2 p} + { b_{0}} \;  
\label{eff-H}
\end{equation}
where ${ M}$ is the neutrino mass matrix in the CPT conserving limit. As mentioned above, the $b_{\mu}$  parameterize the extent of CPT violation.\\

In many experimental situations, the neutrino passes through appreciable amounts of matter.  Accounting for this, the  Hamiltonian with CPT violation in the  flavour basis \footnote{We note that the matrices appearing in the three terms in eq \ref{eff-H} can in principle be diagonal in different bases, one in which the neutrino mass matrix is diagonal, a second one in which the Lorentz and CPT violating interactions are diagonal, and a third flavour basis. } becomes
\begin{equation}
 {H}_{f} = \frac{1}{2E}. {U}_{0}. D(0,\Delta{m}_{21}^{2},\Delta{m}_{31}^{2}). U_{0}^{\dagger} + U_{b}. D_b(0,\delta{b}_{21},\delta{b}_{31}). U_{b}^{\dagger} +  D_m(V_{e},0,0)
\label{Hf}
\end{equation}
where $ U_{0}$ $\&$ $ U_{b}$ are unitary matrices and $V_e$ = $\sqrt{2} G_{F} N_{e}$,  where $G_F$ is the fermi coupling constant and $N_e$ is the electron number density. Value of  $V_e$ = 0.76 $\times$ $10^{-4}$ $\times$ $Y_{e}$ [$\frac{\rho}{g/cc}$] eV, where $Y_{e}$ is the fraction of electron, which is $\approx$ 0.5 for earth matter and $\rho$ is matter density inside earth.   D, $ D_m$ and $ D_b$ are diagonal matrices wih their elements as listed. Here $\delta{b_{i1}} \equiv b_i - b_1$ for $i = 2,3$, where $b_1$, $b_2$ and $b_3$ are the eigenvalues of ${b}$. \\

As is well-known, in standard neutrino oscillations, $ U_{0}$ is parametrized by three mixing angles $(\theta_{12}, \theta_{23}, \theta_{13})$ and one phase $\delta_{CP}$.\footnote{In general for an N dimensional unitary matrix, there are $N$ independent  rotation angles (i.e. real numbers) and  $N(N+1)/2$ imaginary quantities (phases) which define it. For Dirac fields, $(2N-1)$ of these may be absorbed into the representative spinor, while for Majorana fields this can be done for N phases. In the latter case, the $N-1$ additional phases in $ U_{0}$ become irrelevant when the product $ M  M^\dagger$ is taken.}
Similarly (see footnote), any  parametrization of the matrix $ U_{b}$, needs three angles $(\thb_{12}, \thb_{23}, \thb_{13})$ and six phases. Thus $H_{f}$ contains six mixing angles  $(\theta_{12}, \theta_{23}, \theta_{13}, \thb_{12}, \thb_{23}, \thb_{13} )$ and seven phases. \\
 
It is clear that the results will depend on the mixing angles in the CPT violation sector. In what follows, we examine the effects of  three different representative  sets of mixing angles,  1) small mixing ($\thb_{12}=6\,^{\circ}, \thb_{23}=9\,^{\circ}, \thb_{13}=3\,^{\circ}$), 2) large mixing ($\thb_{12}=38\,^{\circ}, \thb_{23}=45\,^{\circ}, \thb_{13}=30\,^{\circ}$) and the third set 3)  uses the same values  as the mixing in neutrino UPMNS,  ($\thb_{12}=\theta_{12}, \thb_{23}=\theta_{23}, \thb_{13}=\theta_{13}$). We use the recent best fit neutrino oscillation parameters in our calculation as mentioned in table 1.

\begin{figure}[!t]
\centering
\subfloat[][Case 1, NH]{
 \includegraphics[width=0.33\textwidth, keepaspectratio]{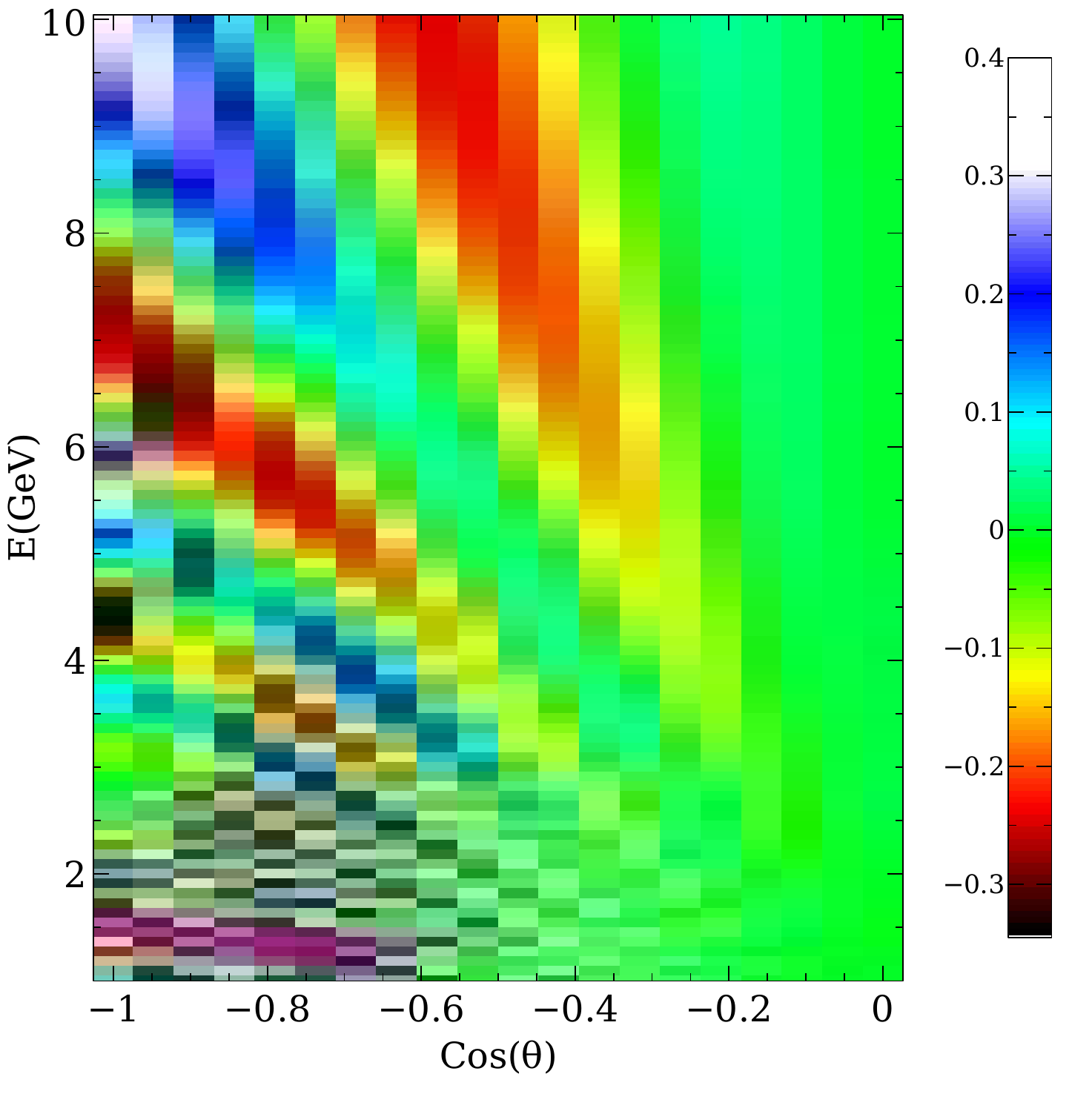}
  \label{fig:Oscgrm_Pmm_smallmix_NH}
}
\subfloat[][Case 1, IH]{
 \includegraphics[width=0.33\textwidth, keepaspectratio]{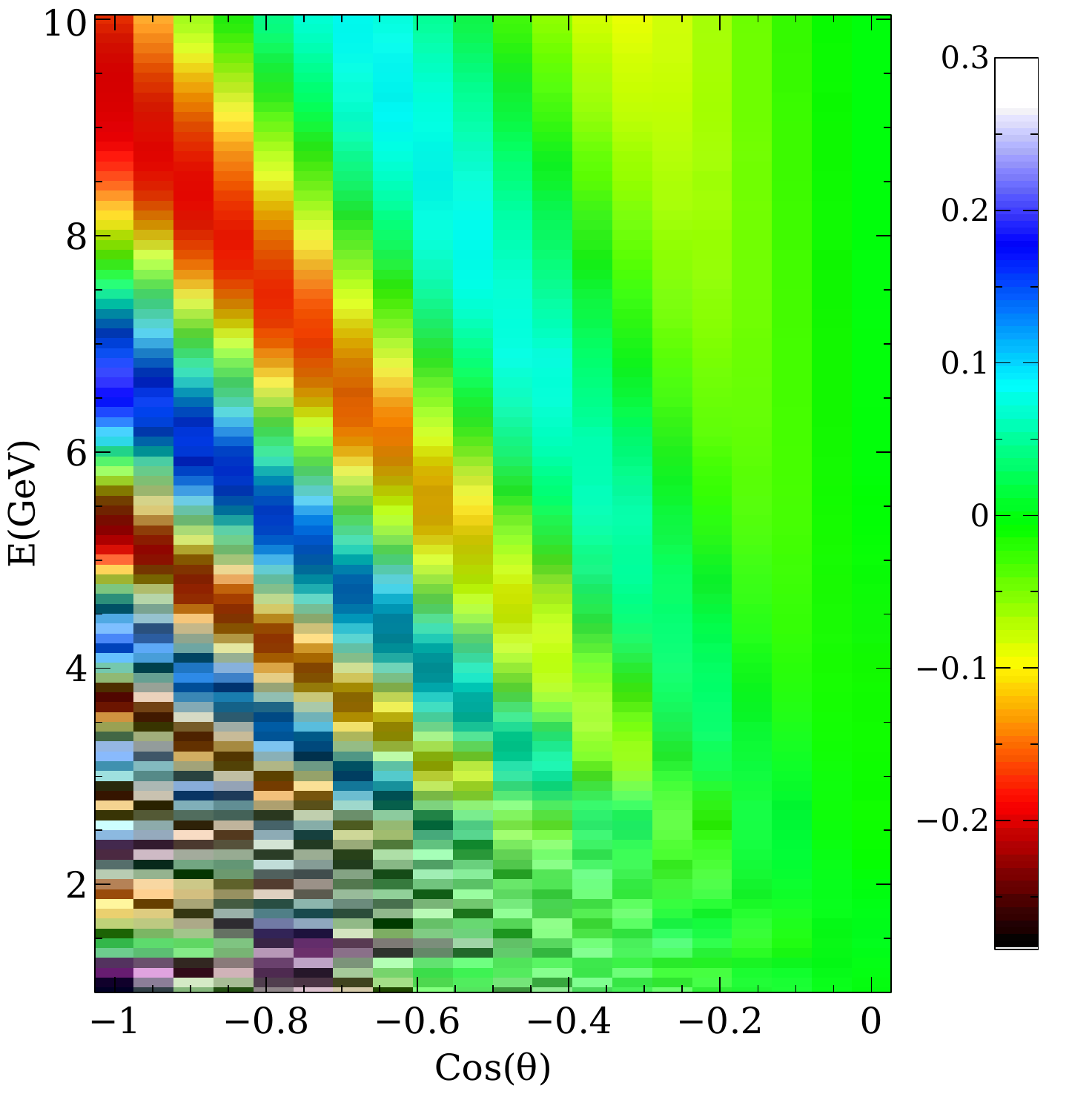}
  \label{fig:Oscgrm_Pmm_smallmix_IH}
}
\quad
\subfloat[][Case 2, NH]{
 \includegraphics[width=0.33\textwidth, keepaspectratio]{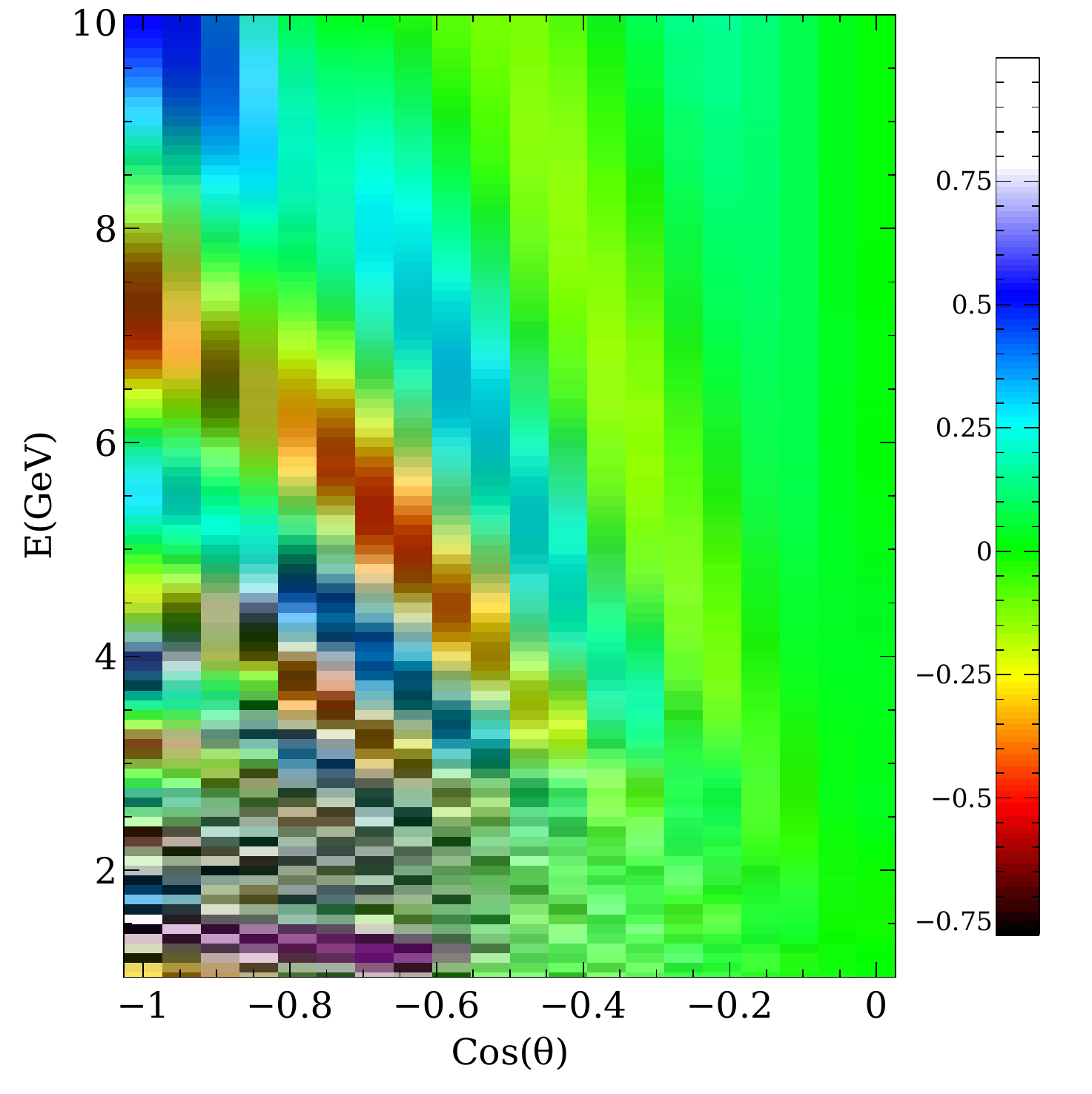}
  \label{fig:Oscgrm_Pmm_largmix_NH}
}
\subfloat[][Case 2, IH]{
 \includegraphics[width=0.33\textwidth, keepaspectratio]{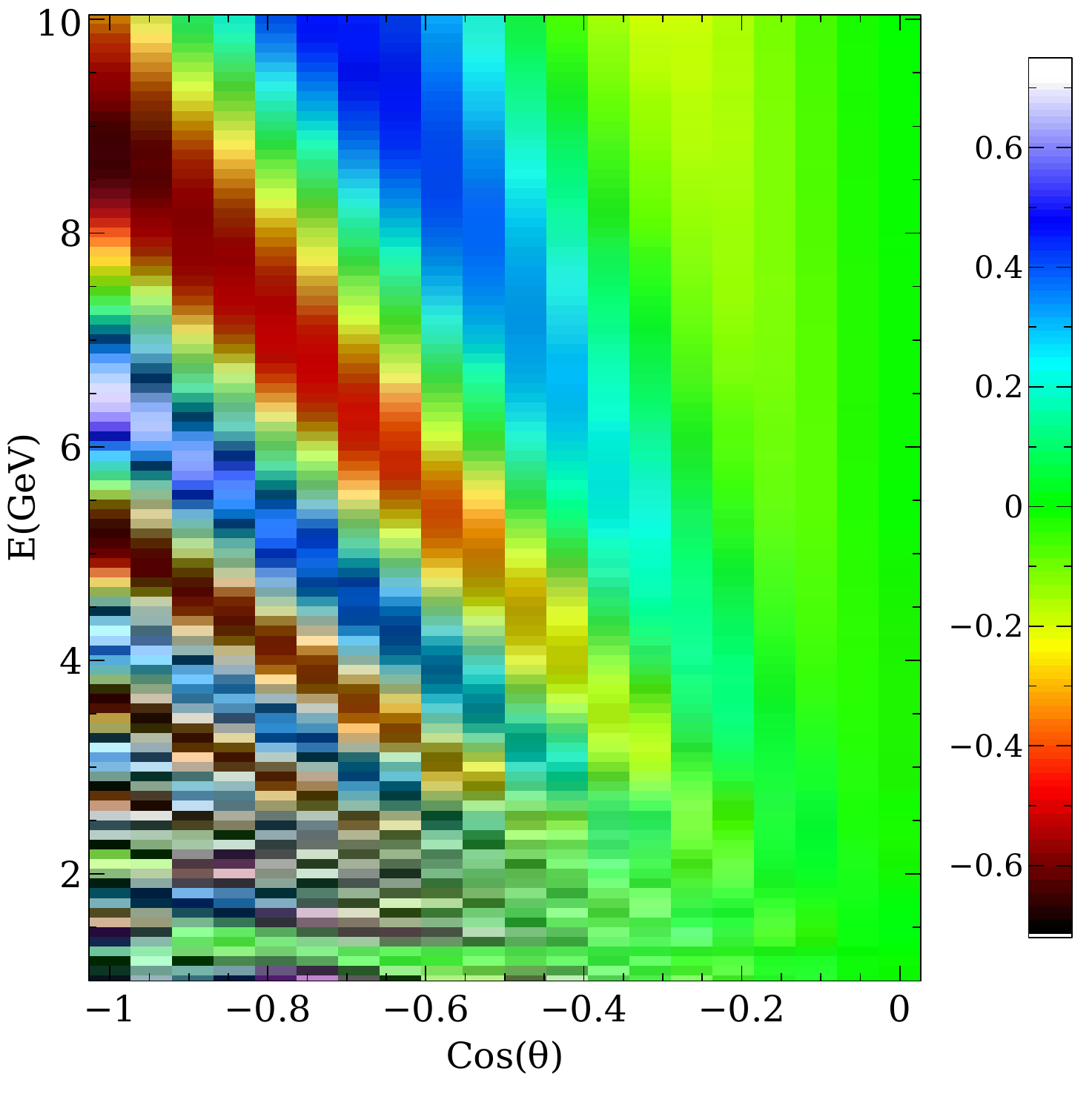}
  \label{fig:Oscgrm_Pmm_larmix_IH}
}
\quad

\subfloat[][Case 3, NH]{
 \includegraphics[width=0.33\textwidth, keepaspectratio]{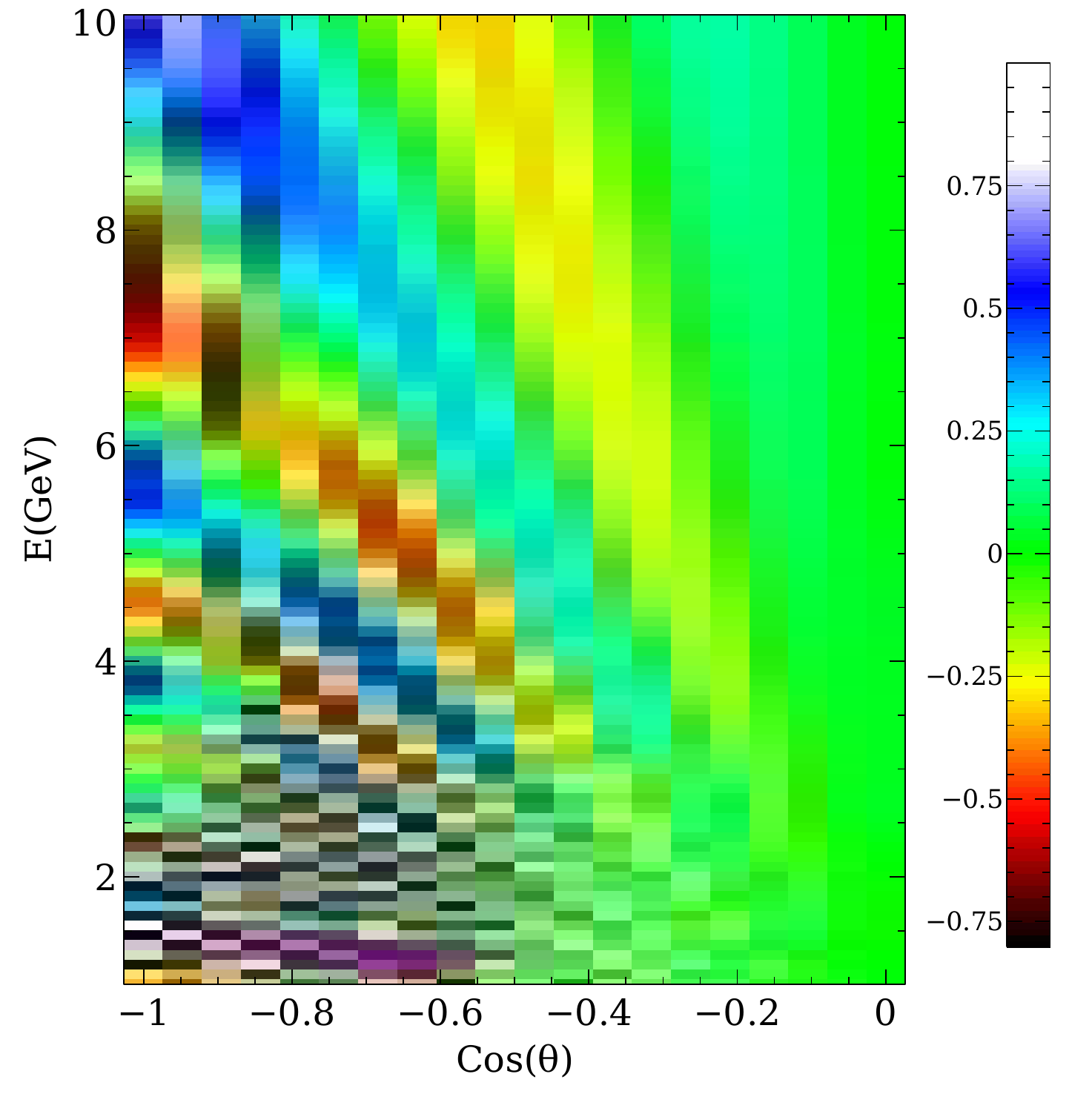}
 
  \label{fig:Osc_Pmm_numix_NH}
}
\subfloat[][Case 3, IH]{
 \includegraphics[width=0.33\textwidth, keepaspectratio]{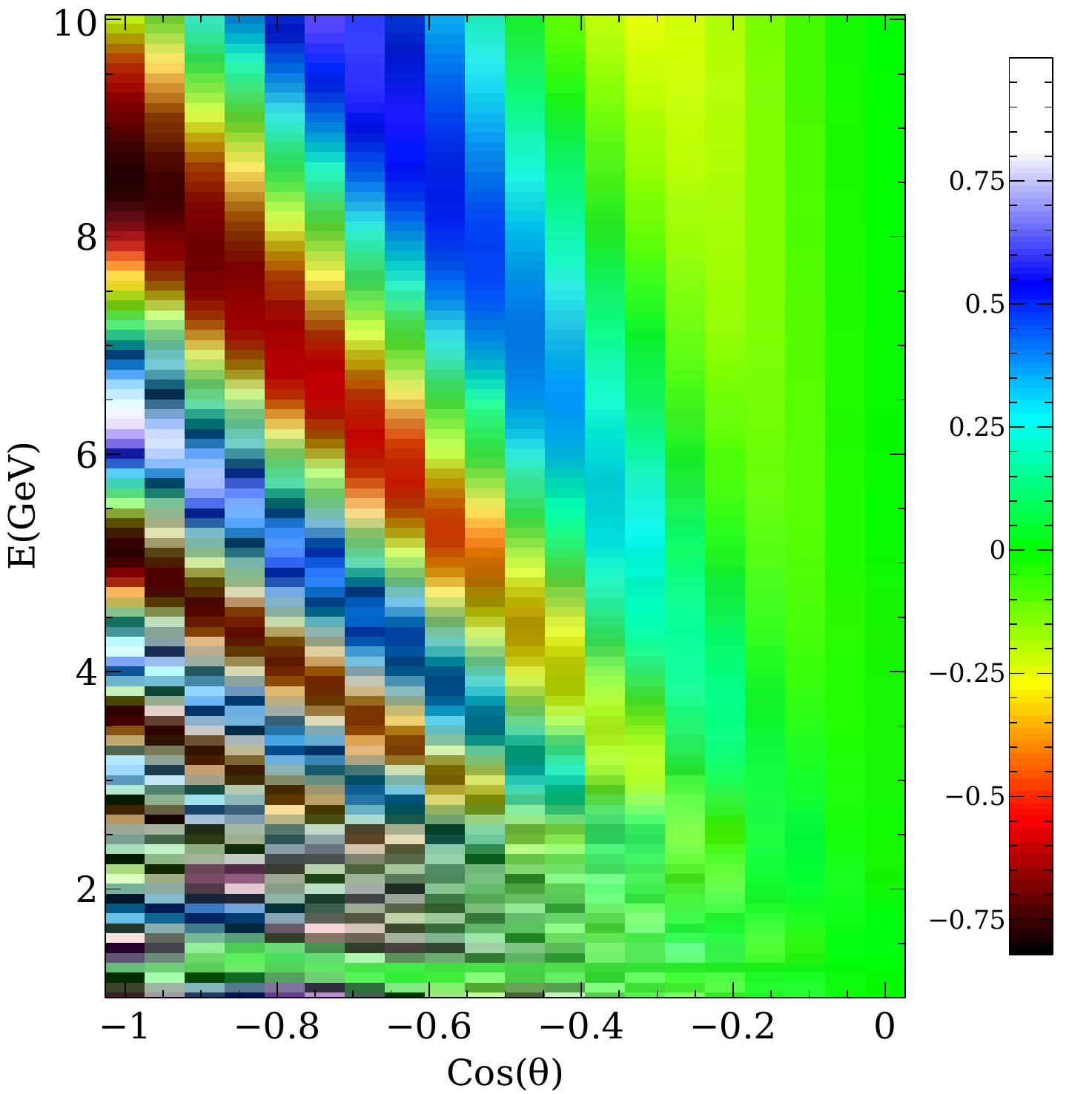}
  \label{fig:Osc_Pmm_numix_IH}
}

\caption{
\label{fig:Osc_Pmm} 
The oscillograms of $\Delta{P} = (P_{\nu_{\mu}\nu_{\mu}}^{U_{b}\neq 0}  - P_{\nu_{\mu}\nu_{\mu}}^{U_{b}= 0}$) for 3 different mixing cases have been shown. The value of  $\delta{b}_{31}=3\times 10^{-23} GeV$ is taken for $U_{b}\neq 0$. Left and right panels are for Normal and Inverted hierarchy respectively.}
\end{figure}

\newpage
Before going into the detailed numerical calculations, we can roughly estimate the bound on CPTV term. As an example, let us assume case 3) $\it{i.e}$ when $U_{b}=U_{0}$, then $\delta{b}$ can effectively be added to $\frac{\Delta{m^{2}}}{2E}$. If we take 10 GeV for a typical neutrino energy, the value of $\frac{\Delta{m^{2}}}{2E}$ will be about $10^{-22}$ GeV. Assuming the CPT violating term to be of the same order, and assuming that neutrino mass splitting can be measured at ICAL to $10\%$ accuracy, we expect sensitivity to  $\delta{b}$ values of approximately  around $10^{-23}$ GeV.

From case 3 above, we note further that in its probability expressions, $\delta{b}_{21}$  will always appear with the  smaller (by a factor of 30) mass squared difference $\Delta{m}^{2}_{21}$. Thus we expect  its effects on oscillations will be subdominant in general,  limiting the capability of atmospheric neutrinos to constrain it, and in our work we have not been able to put useful constraints on $\delta{b}_{21}$. Thus, our effort has been to find a method which will give the  most stringent bounds on CPT violation as parametrized by $\delta{b}_{31}$. For simplicity, all phases are set to zero, hence the distinction between Dirac and Majorana neutrinos with regard to the number of non-trival phases does not play a role in what follows. Moreover, in the approximation where the effects of $\delta{b}_{21}$ are much smaller than those of $\delta{b}_{31}$, the impact of at least some of the nontrivial phases anyway will be negligible.  We also study the effect and impact  of hierarchy in putting constraints on CPTV violating terms.\\

 We also note here previously obtained limits on  the parameters of $U_b$. The solar and KamLAND data \cite{Bahcall} gives the bound $\delta {b}$ $\lesssim$ 1.6 $\times$ $10^{-21}$ GeV. In \cite{Datta} by studying the ratio of total atmospheric muon neutrino survival rates,($\it{i.e}$ two flavour approach different from the one in the present paper), it was shown that, for a 50 kt magnetized iron calorimeter, $\delta {b}$ $\lesssim$ 3 $\times$  $10^{-23}$ GeV should be attainable with a $400$ kT-yr exposure. Using a two-flavor analysis, it was noted in \cite{Barger} that a long baseline (L = 735 km) experiment with a high energy  neutrino factory  can constrain $\delta {b}$ to $\lesssim$ $10^{-23}$ GeV. A formalism for a three flavour analysis was presented in \cite{amol} and bounds of the order of  $\delta {b}$ $\lesssim$ 3 $\times$  $10^{-23}$ GeV were calculated for the upcoming NoVA experiment and for neutrino factories.  It has also been shown in \cite{Samanta}, that a bound of $\delta{b_{31}}$ $\lesssim$ 6 $\times$  $10^{-24}$ GeV at 99$\% $ CL can be obtained with a 1 Mt-yr magnetized iron detector. Global two-flavor analysis of the full atmospheric data and long baseline K2K data puts the bound $\delta {b}$ $\lesssim$ $10^{-23}$ GeV \cite{Gonzalez}.\footnote{Note that the bounds obtained in these papers, and the bounds that we will obtain below,  are on the absolute value of $\delta{b}$, since in principle this quantity can be either positive or negative in the same way the $\Delta{m}^{2}_{ij}$ can be positive or negative. In our plots, where necessary, we assume it to be positive for simplicity.}\\

\begin{figure}[!t]
\centering
\subfloat[][Case 1, NH]{
 \includegraphics[width=0.33\textwidth]{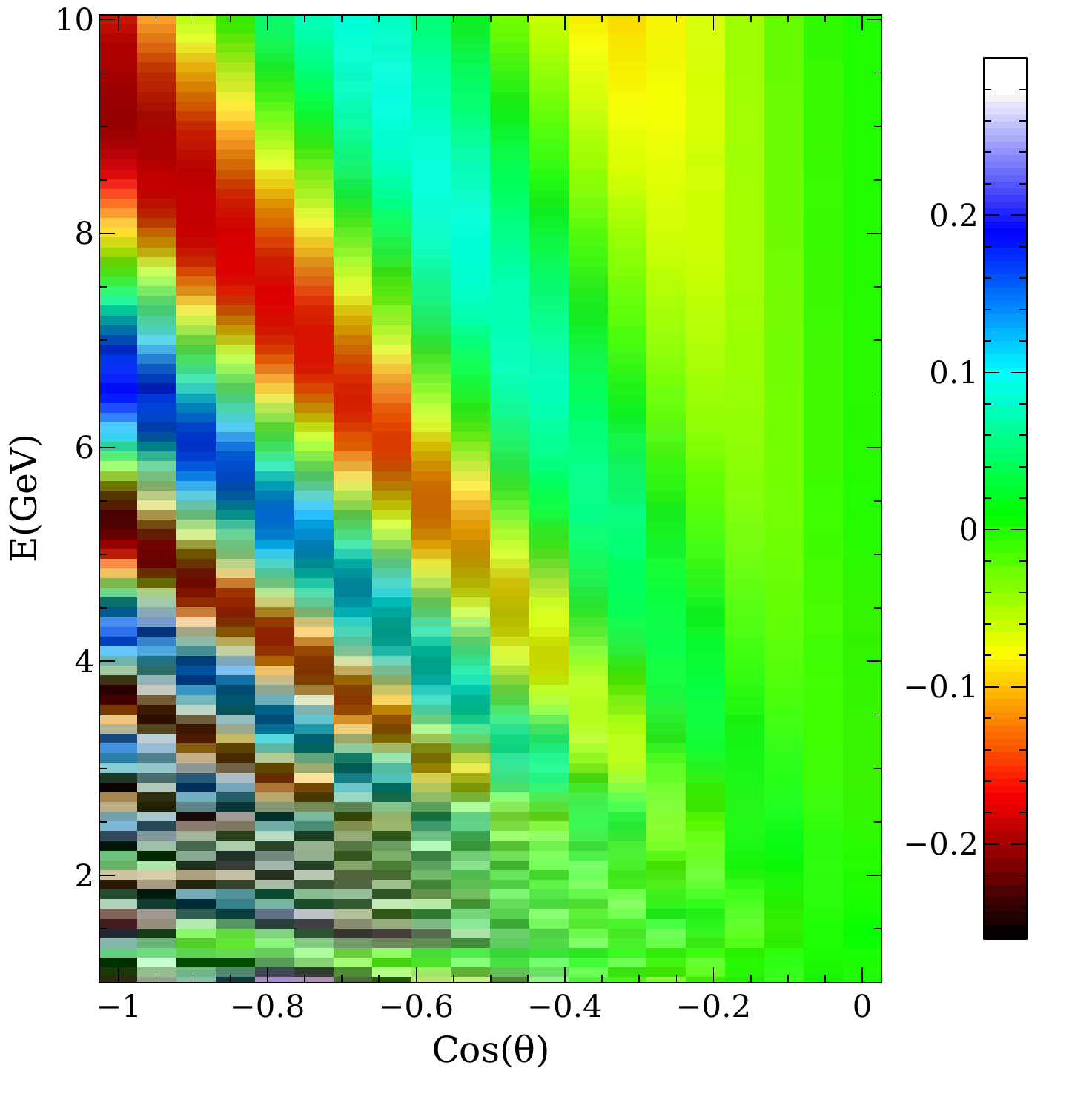}
  \label{fig:Oscgrm_Pmmbar_lowmix_NH}
}
\subfloat[][Case 1, IH]{
 \includegraphics[width=0.33\textwidth]{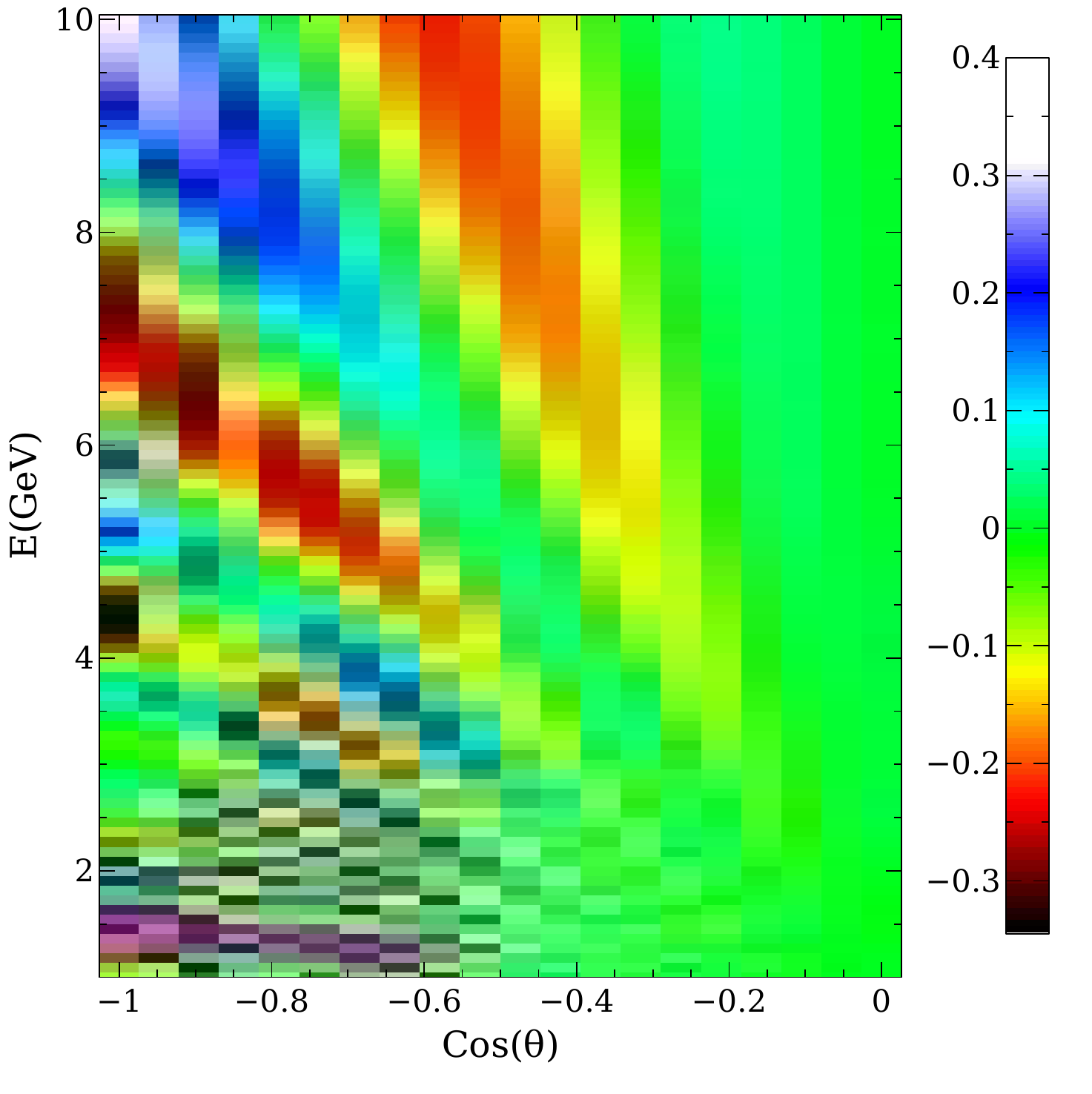}
  \label{fig:Oscgrm_Pmmbar_lowmix_IH}
}
\quad

\subfloat[][Case 2, NH]{
 \includegraphics[width=0.33\textwidth]{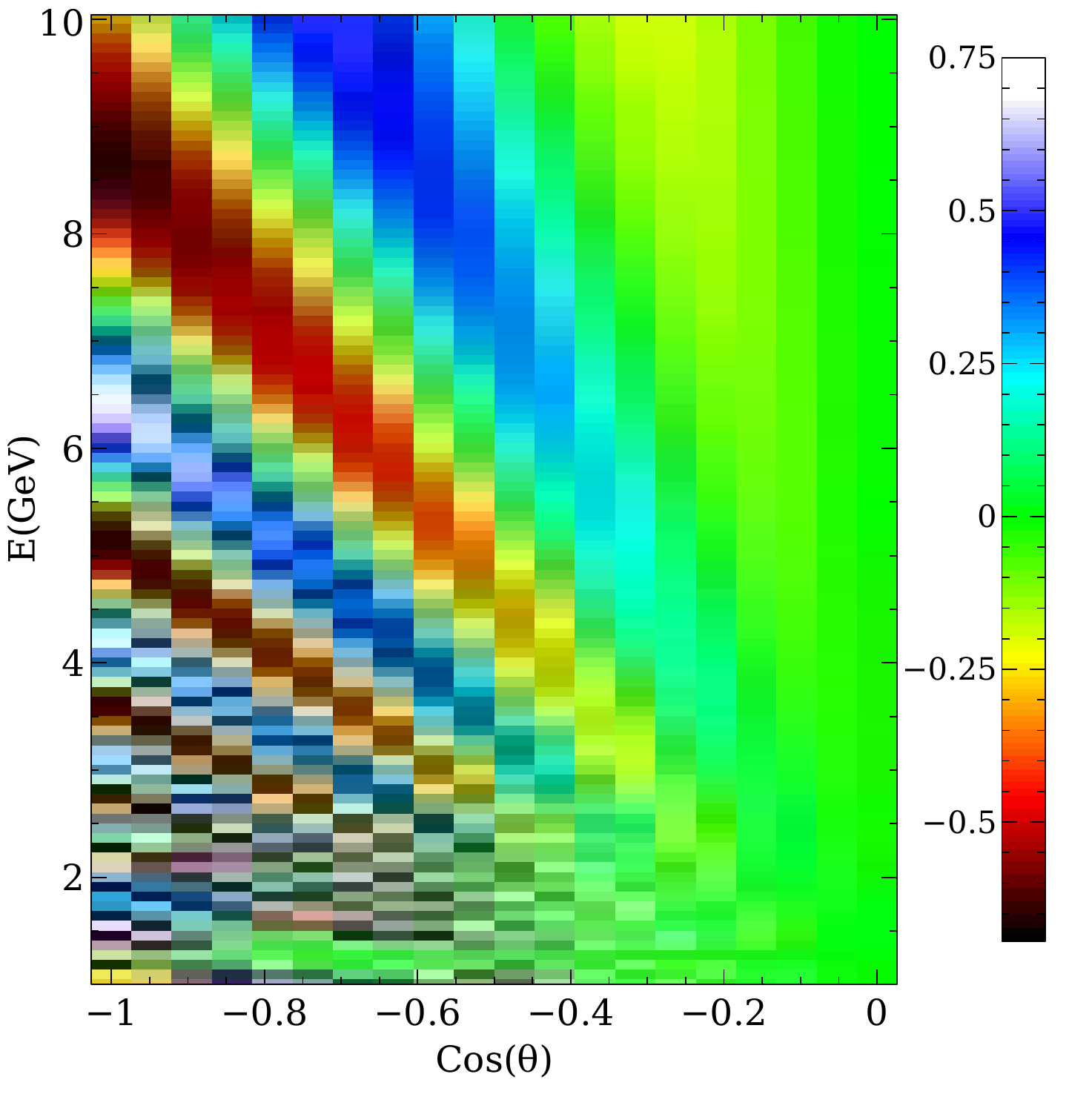}
  \label{fig:Oscgrm_Pmmbar_large_NH}
}
\subfloat[][Case 2, IH]{
 \includegraphics[width=0.33\textwidth]{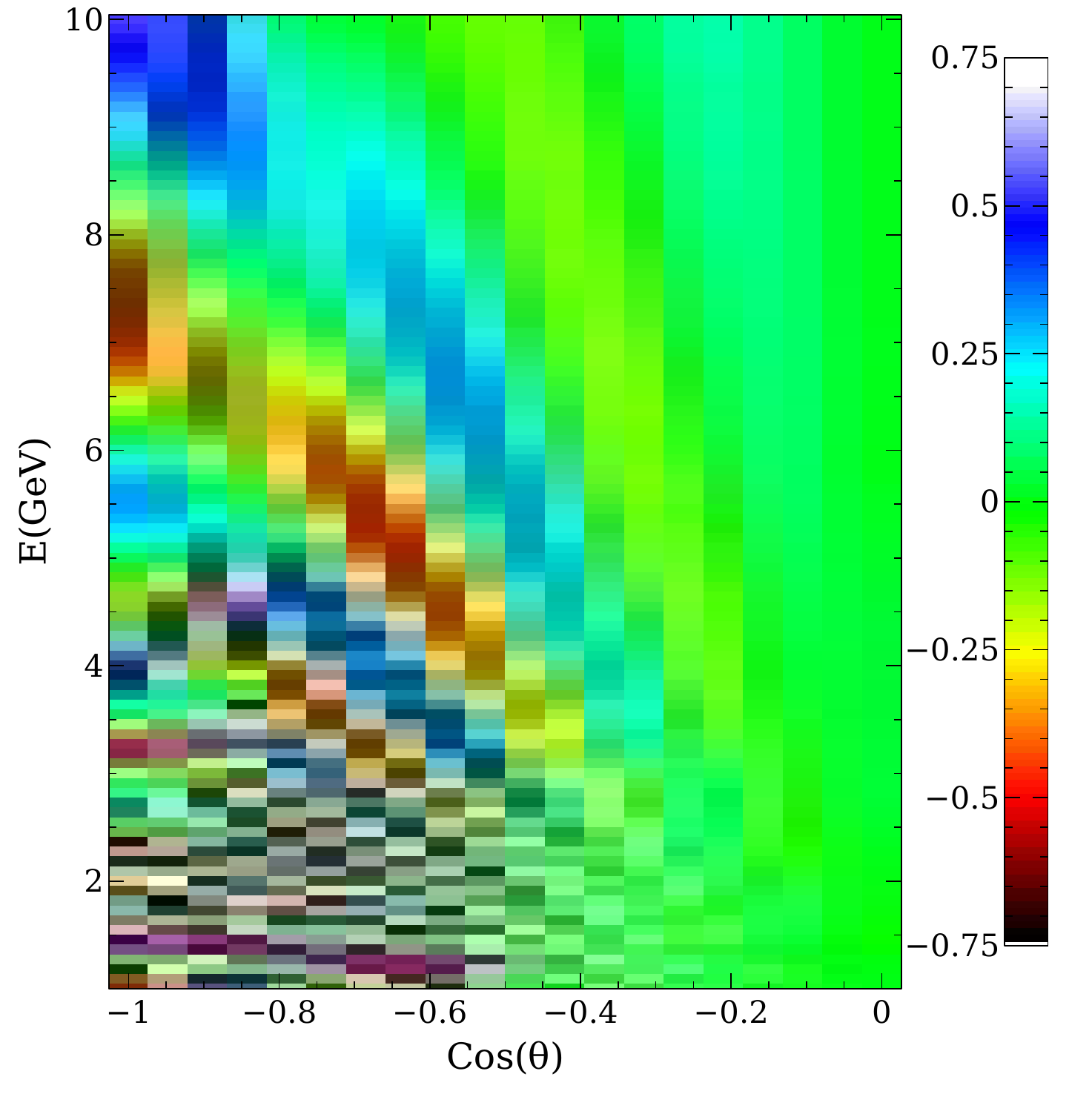}
  \label{fig:Oscgrm_Pmmbar_large_IH}
}
\quad

\subfloat[][Case 3, NH]{
 \includegraphics[width=0.33\textwidth]{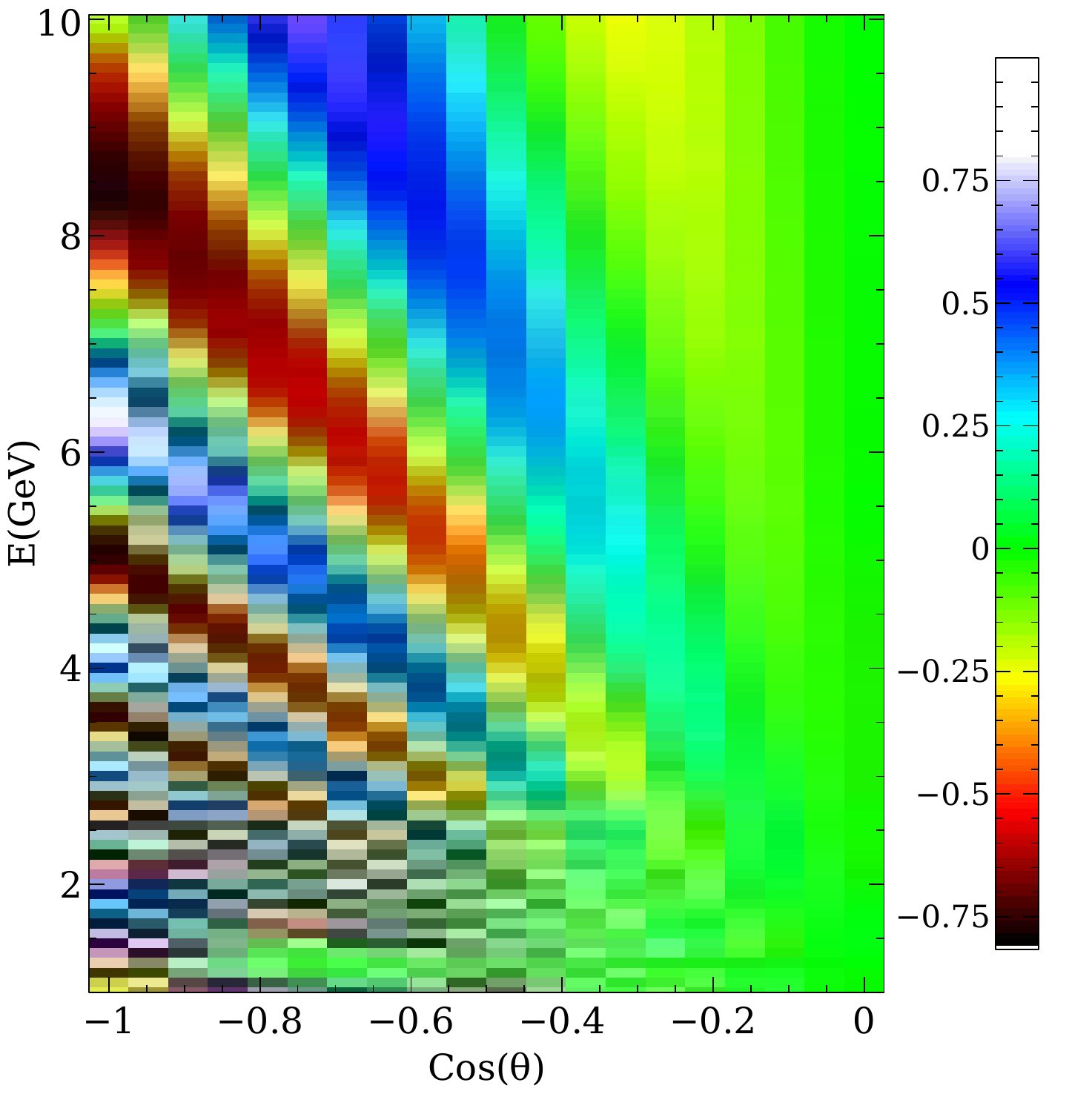}
  \label{fig:Oscgrm_Pmmbar_numix_NH}
}
\subfloat[][Case 3, IH]{
 \includegraphics[width=0.33\textwidth]{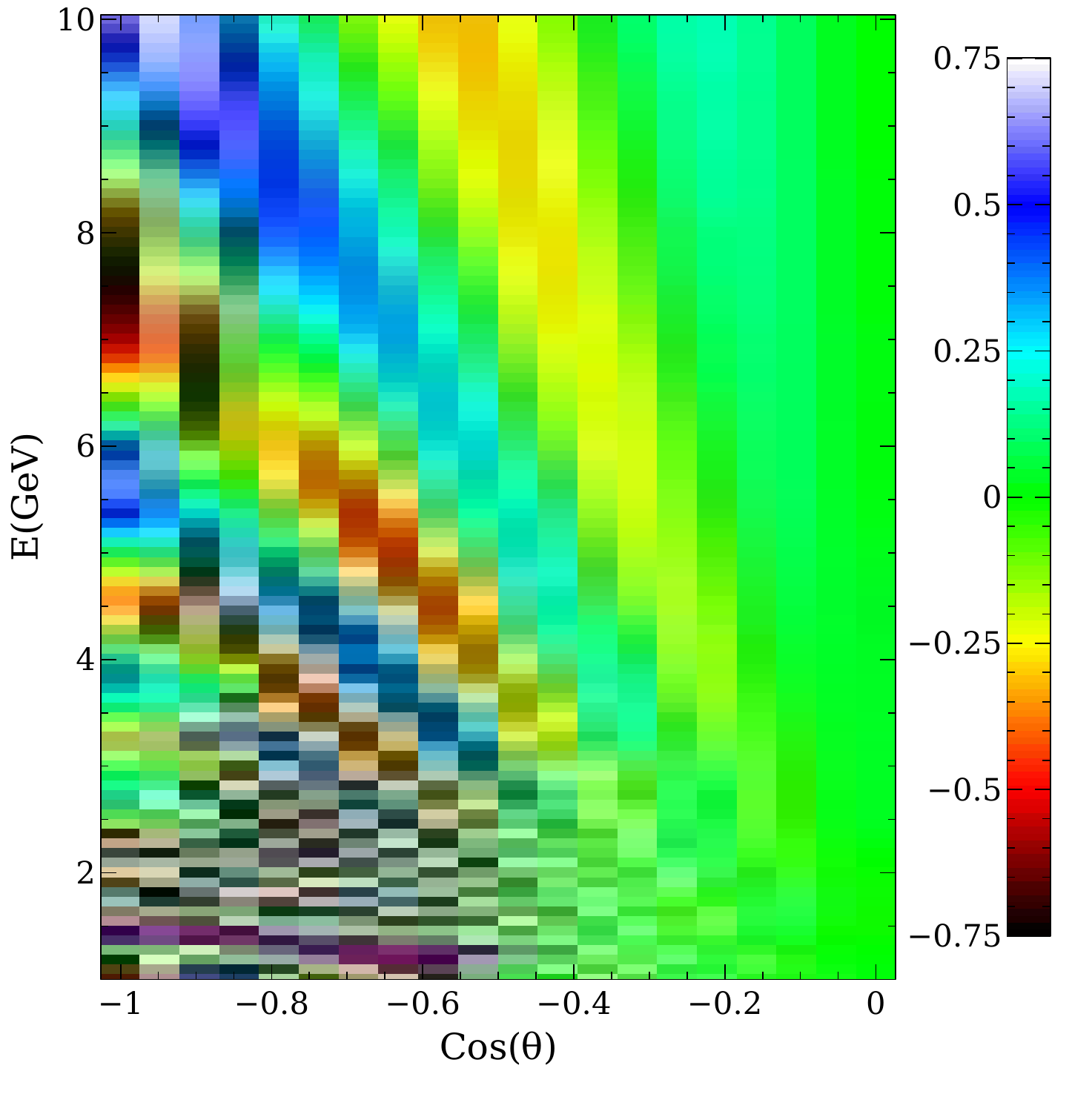}
  \label{fig:Oscgrm_Pmmbar_numix_IH}
}

\caption{
\label{fig:Osc_Pmmbar} 
The oscillograms of $\Delta{P} = (P_{\bar{\nu_{\mu}}\bar{\nu_{\mu}}}^{U_{b}\neq 0}  - P_{\bar{\nu_{\mu}}\bar{\nu_{\mu}}}^{U_{b}= 0}$) for 3 different mixing cases have been shown. The value of  $\delta{b}_{31}=3\times 10^{-23} GeV$ is taken for $U_{b}\neq 0$. Left and right panels are for Normal and Inverted hierarchy respectively.}
\end{figure}

 Prior to discussing the results of our numerical simulations, it is useful to examine these effects at the level of probabilities. We note that the matter target in our case is CP asymmetric, which will automatically lead to effects similar to those induced by $\bf U_{b}$. In order to separate effects arises due to dynamical CPT violation from those originating due to the CP asymmtry of the earth, it helps to consider the 
 difference in the disappearance probabilities with $\bf U_{b}$ effects turned on and off, respectively. We use the difference in probabilities 
 
\begin{equation} 
\Delta{P} = P_{\nu_{\mu}\nu_{\mu}}^{U_{b}\neq 0}  - P_{\nu_{\mu}\nu_{\mu}}^{U_{b}= 0}. 
\end{equation}
 
 We do this separately for  ${\nu_{\mu}}$  and ${\bar{\nu_{\mu}}}$ events with NH and IH assumed as the  true hierarchy. The results are shown  in the figures $1-2$. ( We note that at the event level, the  total muon events receive contributions from both the $P_{\nu_{\mu}\nu_{\mu}}$ disappearace and $P_{\nu_{e}\nu_{\mu}}$ appearance channels, and the same is true for anti-neutrinos. In our final numerical results, we have taken both these contributions into account).

Several general features are apparent in figures 1-2. First, effects are uniformly small at shorter baselines irrespective of the value of the energy. From the 2 flavour analysis, $\it{e.g.}$ \cite{Datta} we recall that the survival probability difference  in vaccuum is proportional to $\sin(\frac{\delta{m}^{2} L}{2E})\sin(\delta{b} L)$. The qualitative feature that CPT effects are larger at long baselines continues to be manifest even when  one incorporates three flavour mixing and the presence of matter, and this is brought out in all the figures.

\begin{figure*}[h]
\begin{center}
\includegraphics[width=0.5 \textwidth]{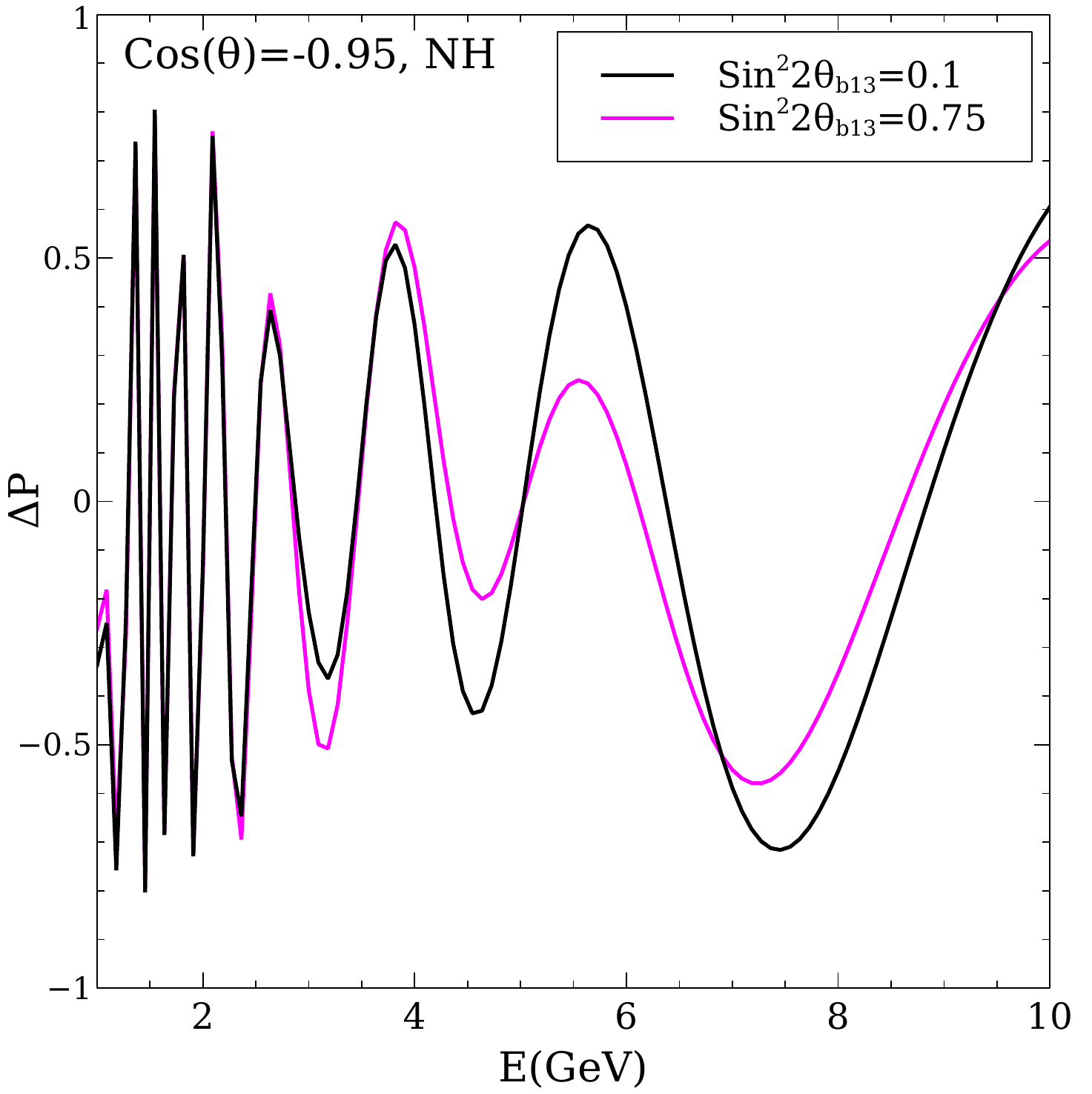}
\vspace{-0.2in}
\caption{\small{ The difference of the $P_{\nu_{\mu}\nu_{\mu}}$ with and without CPTV for $\delta{b}_{31} = 3\times10^{-23} GeV$ for two $\theta_{{b}_{13}}$ values as a function of energy for a specific value of zenith angle. All other oscillation parameters are same.}}
\label{Diffthb13_Plot}
\end{center}
\end{figure*}

Secondly, as is well-known, matter effects are large and resonant for neutrinos and NH, and for anti-neutrinos with IH. Thus in both these cases, they mask the (smaller) effect of CPT stemming from $ U_{b}$. Hence for neutrino events, CPT sensitivity is significantly higher if the hierarchy is inverted as opposed to normal, and the converse is true for anti-neutrino events.
Finally, effects are largest for cases 3) and 2), and smaller for case 1). The effect is smaller for case 1) is due to the fact that mixing is very small compared to other two. The origin of the difference for the case 2) and 3) is likely due to the fact that CPT violating effects are smaller when $\theta_{b13}$ is large, as shown in fig \ref{Diffthb13_Plot}.  \\

We carry through this mode of looking at the difference between the case when $\bf U_{b}$ is non-zero and zero repectively to the event and $\chi^{2}$ levels in our calculations below. To use the lepton charge identification capability of a magnetized iron calorimeter optimally,  we calculate $\chi^{2}$ from $\mu^{-}$ and $\mu^{+}$ events separately.  Following this procedure,  the  contribution arising through matter being CPT asymmetric will expectedly cancel for any  given zenith angle and energy. The numerical procedure adopted and the details of our calculation are provided in the following section.

\section{Numerical procedure}

  Our work uses the   ICAL detector as a reference configuration, but the qualitative content of the results will hold for any similar detector. Magnetized iron calorimeters typically have  very good energy and direction resolution for reconstructing $\mu^+$ and $\mu^-$ events.  The analysis proceeds in two steps: (1) Event simulation  (2) Statistical procedure and the  $\chi^2$ analysis.
\subsection{Event simulation}

 We use the NUANCE\cite{nuance} neutrino generator to generate  events. The  ICAL detector composition and geometry are incorporated within NUANCE and atmospheric neutrino fluxes( Honda et al. \cite{Honda}) have been used. In order to reduce the Monte Carlo (MC) fluctuations in the number of events given by NUANCE, we generate a very large number of neutrino events with an exposure of 50 kt $\times$ 1000 years and then finally normalize to 500 kt-yr.\\

Each changed-current neutrino event is characterized by neutrino energy and neutrino zenith angle, as well as by a muon energy and muon zenith angle. In order to save on computational time,  we  use a re-weighting algorithm to generate oscillated events. This algorithm, takes the  neutrino energy and angle for each event and calculates probabilities $P_{\nu_{\mu} \nu_{\mu}}$ and $P_{\nu_{\mu}\nu_{e}}$  for any given set of oscillation parameters. It then  compares it with a random number r between 0 to 1. If $r < P_{\nu_{\mu}\nu_{e}}$, then it is classified as a $\nu_{e}$ event. If $ r >  (P_{\nu_{\mu}\nu_{e}} + P_{\nu_{\mu} \nu_{\mu}})$, it classified as a $\nu_{\tau}$ event. If $P_{\nu_{\mu}\nu_{e}} \leq r \leq (P_{\nu_{\mu}\nu_{e}} + P_{\nu_{\mu} \nu_{\mu}})$, then it is considered to  come from an atmospheric $\nu_\mu$ which has survived as a $\nu_\mu$. Similarly muon neutrinos from the oscillation of $\nu_{e}$ to $\nu_{\mu}$ are also calculated using this  reweighting method.\\

Oscillated muon events are  binned as a function of muon energy and muon zenith angle.  We have divided each of the ten  energy bins into 40 zenith angle bins.  These binned data are folded with detector efficiencies and resolution functions as described in equation (6) to simulate reconstructed muon events. In this work we have used the (i) muon reconstruction efficiency (ii) muon charge identification efficiency (iii) muon energy resolution (iv) muon zenith angle resolution, obtained by the INO collaboration\cite{muon_resolution}, separately for $\mu^{+}$ and $\mu^{-}$.\\

The measured muon events after implementing efficiencies and resolution are  
\begin{equation}
N(\mu^{-})=\int \mathrm{d}E_{\mu} \int \mathrm{d}\theta_{\mu}[ R_{E_{\mu}}  R_{\theta_{\mu}}(R_{eff}C_{eff}N_{osc}(\mu^{-})+\bar{R}_{eff}(1-\bar{C}_{eff})N_{osc}(\mu^{+}))] 
\label{event}
\end{equation}

where $R_{eff},C_{eff}, \bar{R}_{eff}, \bar{C}_{eff}$ are reconstruction and charge identification efficiencies for $\mu^{-}$ and $\mu^{+}$ respectively, $N_{osc}$ is the number of oscillated muons in each true muon energy and zenith angle bin and $R_{E_{\mu}}$, $R_{\theta_{\mu}}$ are energy and angular resolution functions. \\
The energy and angular resolution function in Gaussian form are given by
\begin{equation}
R_{E}=\frac{1}{\sqrt{2\pi}\sigma_{E}}\exp \bigg[-\frac{(E_m-E_t)^2}{2\sigma_{E}^2}\bigg]
\end{equation}

\begin{equation}
 R_{\theta}=N_{\theta} \exp \bigg[-\frac{(\theta_t-\theta_m)^2}{2(\sigma_{\theta})^2} \bigg]
 \end{equation}
 Here $E_{m}$, $E_{t}$ and $\theta_{m}$,$\theta_{t}$ are measured and true energy and angle respectively.  $N_{\theta}$ is the normalization constant, $\sigma_{\theta},\sigma_{E}$ are angular and energy smearing of muons. $\sigma_{\theta},\sigma_{E}$ are obtained from ICAL simulations \cite{muon_resolution}.\\
 \subsection{Statistical procedure and the $ \chi^2$ analysis}

 \begin{table}[H]

  \begin{center}
  \begin{tabular}{|c|c|c|c|c|}
     
  \hline
  Oscillation parameter & Best fit values &  Oscillation parameter & Best fit values \\
  \hline
  $\sin^2 2\theta_{12}$ & 0.86 & $\delta_{\text{CP}}$ & 0.0 \\
  \hline
  $\sin^2 2\theta_{23}$ & 1.0 & $\sin^2 2\theta_{b12}$ & 1) 0.043, 2) 0.94, 3) 0.86\\
			 
  \hline
  $\sin^2 2\theta_{13}$ & 0.1 & $\sin^2 2\theta_{b23}$ & 1) 0.095, 2) 1.0, 3) 1.0 \\
			 
  \hline
  $\Delta m_{21}^2 \: (\rm eV^2)$  & 7.5 $\times \: 10^{-5}$ & $\sin^2 2\theta_{b13}$  & 1) 0.011, 2) 0.75, 3) 0.1\\
  \hline
  $|\Delta m_{32}^2| \: (\rm eV^2)$ & 2.4 $\times \: 10^{-3}$ & $\delta_b, \phi_{b2}, \phi_{b3}$ & 0.0, 0.0, 0.0 \\
					 
  \hline
  
  \end{tabular}
  \end{center}

  \caption{True values of the  oscillation parameters used in the analysis}
  \label{tab_osc_param_input}

  \end{table}

 \begin{table}[H]

  \begin{center}
  \begin{tabular}{|c|c|c|c|c|}
     
  \hline
  Uncertainties & Values \\
  \hline
  Flux Normalization & $20\%$ \\
  \hline
  Tilt Factor & $5\%$\\
			 
  \hline
  Zenith angle dependence & $5\%$\\
			 
  \hline
  Overall cross section  & $10\%$\\
  \hline
  Overall systematic & $5\%$ \\
					 
  \hline
  
  \end{tabular}
  \end{center}

  \caption{Systematic uncertainties used in the $\chi^{2}$ analysis}
  \label{sys}

  \end{table}

 We have generated event rate data with the true values of oscillation parameters given in Table 1 and assuming no CPT violation, these are  defined as $N^{ex}$. They are then fitted with another set of data, labelled as ($N^{th}$), where we have assumed CPT violation. The statistical significance of the difference  between these two sets of data will provide constraints on the  CPT violation parameters.\\
We define $\chi^{2}$ for the data as
\begin{equation}
\chi^2_{pull}=min_{\xi_{k}} [2(N^{th'}-N^{ex}-N^{ex}\ln(\frac{N^{th'}}{N^{ex}})) +\sum_{k=1}^{npull}\xi_{k}^2]
\end{equation}
where
\begin{equation}
N^{th'}=N^{th}+\sum_{k=1}^{npull}c^{k}\xi_k
\end{equation}
 npull is the number of pull variable, in our analysis we have taken npull=5. $\xi_{k}$ is the pull variable and $c^{k}$ are the systematic uncertainties. We have used  5 systematic uncertainties in this analysis as mentioned in Table 2 as generally used in the other analysis of the collaboration. We have assumed a Poissonian distribution for $\chi^{2}$ because for higher energy bins the  number of atmospheric events  will be  small.\\

Since ICAL can discriminate charge of the particle, it is useful to  calculate $\chi^{2}(\mu^{-})$ and $\chi^{2}(\mu^{+})$ separately
for $\mu^{-}$ and $\mu^{+}$ events and then added to get total $\chi^{2}$. We have marginalized the total $\chi^{2}$ within a $3\sigma$ range of the best fit value. $\chi^2$ has been marginalized 
 over the oscillation parameters $\Delta{m}_{31}^{2}$, $\theta_{23}$, $\theta_{13}$, $\delta_{CP}$, $\delta{b_{21}}$ for both normal and inverted hierarchy with $\mu^{+}$ and $\mu^{-}$ separately for given set of input data. \\

\section{Results}

\begin{figure}[t!]
\centering
\subfloat[]{
 \includegraphics[width=0.4\textwidth]{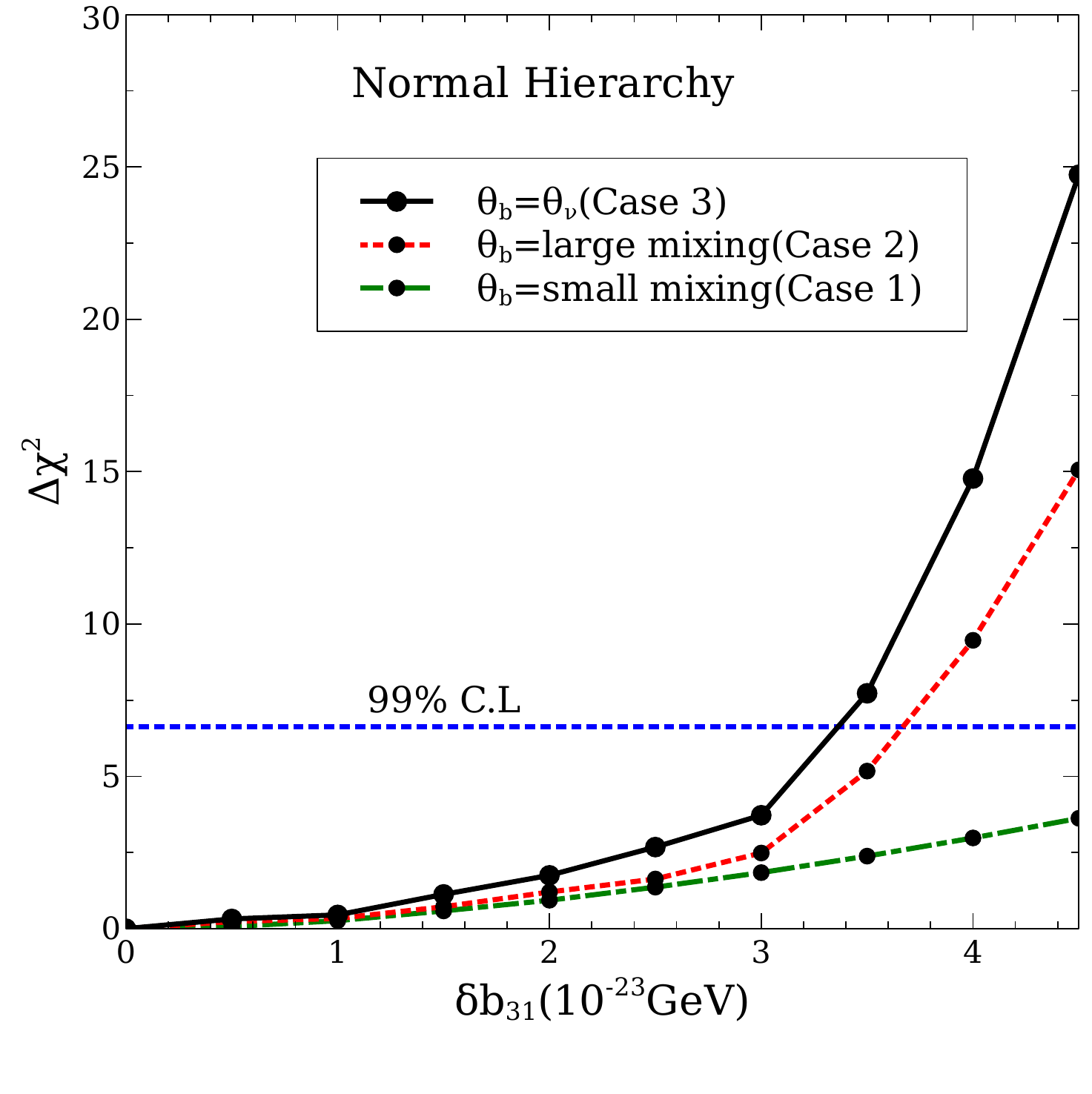}
  
}
\subfloat[]{
 \includegraphics[width=0.4\textwidth]{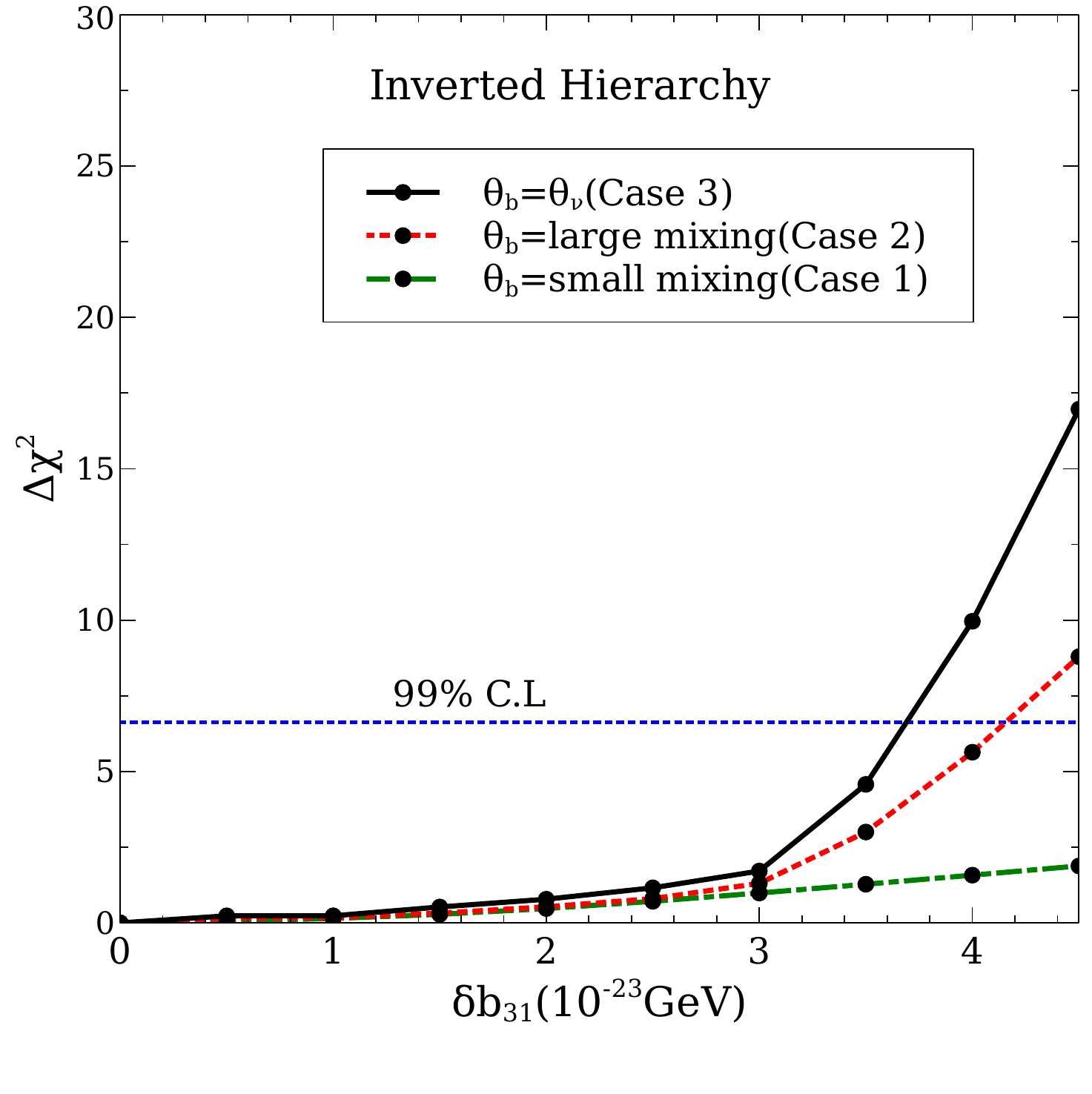}
  
}

\caption{
\label{NH_Plot} 
$\Delta{\chi^{2}}$ as a function of $\delta{b}_{31}$ for different mixing of $\theta_{b}$. Best fit oscillation parameters used as mentioned in table 1. The results are marginalized over  $\theta_{23}$, $\theta_{13}$, $\delta_{CP}$, $\Delta{m}_{31}^{2}$ and $\delta{b_{21}}$. Left and right panel are for Normal and Inverted hierarchy respectively. }
\end{figure}

\begin{figure*}[h]
\begin{center}
\includegraphics[width=0.5 \textwidth]{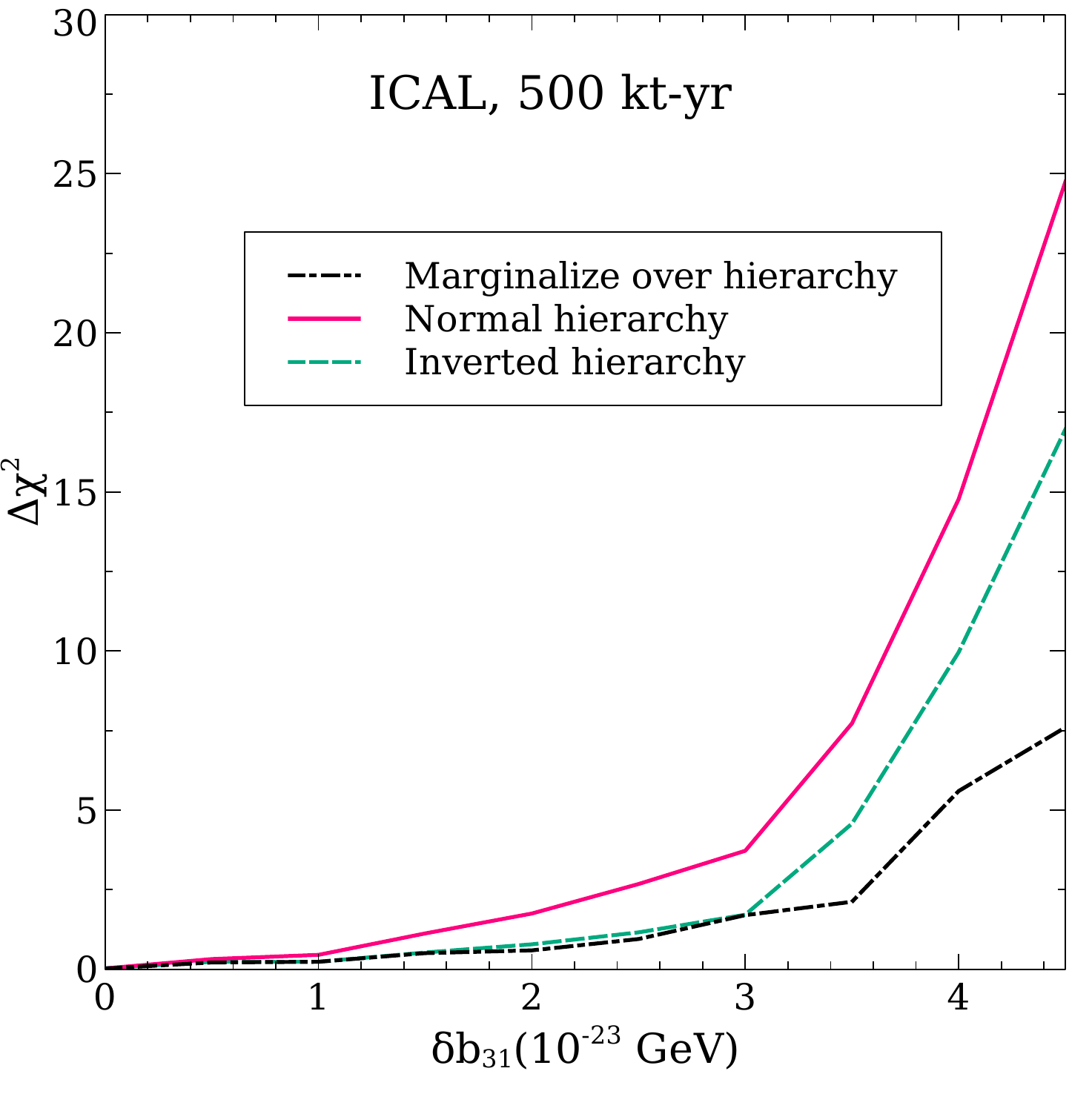}
\vspace{-0.2in}
\caption{\small{$\Delta{\chi^{2}}$ as a function of $\delta{b}_{31}$ for case 3) is shown. The Red curve represents that both sets of data has Normal hierarchy as true hierarchy, and the green curve for Inverted hierarchy respectively. Black curve shows the bounds where hierarchy is marginalized over.}}
\label{margin_Plot}
\end{center}
\end{figure*}

Figure(\ref{NH_Plot}) illustrates $\Delta{\chi^{2}}$ analysis performed by marginalizing over all the oscillation parameters $\Delta{m}_{31}^{2}$, $\theta_{23}$, $\theta_{13}$ within a $3\sigma$ range of their best fit values as given in Table 1. $\delta_{CP}$ is marginalized over 0 to $2\pi$. $\delta{b}_{21}$ is marginalized over the range 0 to $5\times10^{-23}$ GeV. Left and right panel are for the  Normal and Inverted Hierarchy respectively. While from figure(\ref{NH_Plot}) we see that  the best  bounds arise for case 3) for both the hierarchy,  where mixing in the CPTV sector is the same as in the case of neutrino mixing, good bounds are also obtainable for large mixing.
Since $\theta_{b12}$ and $\theta_{b23}$ in both cases 2) and 3) are large, the origin of this difference is likely due to the fact that CPT violating effects are smaller when $\theta_{b13}$ is large, as shown in fig \ref{Diffthb13_Plot}.\\ 

 From figure(\ref{NH_Plot}) we see that  $99\%$ C.L. or better constraints on the CPT violating parameter, $\delta{b}_{31}$  are possible for both hierarchies if it is   $\gtrsim$ $4\times10^{-23}$ GeV, if the mixing in the CPTV sector is not small. 

 It is clear from the fig \ref{margin_Plot} that if marginalization over the hierarchy is carried out, the constraints are considerably weaker. Hence a knowledge of the hierarchy  certainly helps in getting useful constraints on CPT. The sensitivities  obtained are comparable  to those anticipated  from other types of experiments and estimates in the literature \cite{Barger,Bahcall,Datta,amol,Samanta}. 
 
\section{Conclusions:}

A magnetized iron calorimeter like ICAL,
with its attributes of  good energy and angular resolution for muons and  its charge identification capability can be a useful tool in investigating Lorentz and CPT violation stemming from physics at higher energy scales, even though its  main physics objective may be  hierarchy determination. Using  resolutions, efficiencies, errors and uncertainties generated by ICAL detector simulation, we have calculated reliable sensitivites to the presence of CPTV in such a detector.\\

 It is, of course, clear that the exact value of the constraint on CPT violation depends on the choice of mixing angles in $U_{b}$. We have carried out our calculations for three representative cases, those for which the mixing is 1) small, 2) large and 3) similar to the PMNS mixing. We find that for both types of hierarchy, ICAL should be sensitive to  $\delta{b}_{31}$ $\gtrsim$ $4\times10^{-23}$ GeV at 99$\%$ C.L. with 500 kt-yr of exposure, unless the mixing in the CPTV sector is small. As discussed earlier, CP (and CPT) effects due to earth matter asymmetry are subtracted out.\\

Our study pertains to the type of CPTV that may be parametrized by Equation 1, which stems from explicit  Lorentz violation, and to the muon detection channel.  We have not considered the CPTV that arises from differing masses for particles and anti-particles. Finally, we note that in order to obtain good sensitivity to CPTV, knowing the hierarchy will be an important asset, which will anyway be done in a detector like ICAL.

\section{Acknowledgments}
We thank the INO physics and simulation group for their help. We also especially thank Amitava Raychaudhuri, Amol Dighe, M. V. N. Murthy and S Uma Sankar for the useful discussions and comments. A.C thanks Pomita Ghoshal and Anushree Ghosh for the help during the work.

\end{document}